# Cosmology analogy for perfect hyperlens

Tao Hou[1,2], Wen Xiao[1,2], Huanyang Chen[1*]

[1]*Department of Physics, Xiamen University; Xiamen, 361005, China*
[2]*Jiujiang Research Institute of Xiamen University; Jiujiang, 332000, China*
*\*kenyon@xmu.edu.cn*

**Abstract:** With the emergence of super-resolution lenses such as superlens and hyperlens, coupled with advancements in metamaterials, the diffraction limit of approximately half wavelength is no longer unbreakable. However, superlenses are easily affected by weak intrinsic losses and hyperlenses cannot achieve perfect imaging, significantly constraining their practical utility. To address these challenges, here we propose a perfect hyperlens based on the metric of de Sitter spacetime in cosmology. Importantly, perfect hyperlens is capable of self-focusing in geometrical optics while supporting propagating waves with exceptionally large wavenumbers, which endows it with key advantages such as ultra-high resolution, no aberration and strong robustness. Furthermore, we demonstrate the hyperbolic focusing performance and mimic the de Sitter spacetime in naturally in-plane hyperbolic polaritons of α–$MoO_3$ films numerically, which can be achieved with a gradient thickness profile. Our work provides cosmological insights into the field regulation in hyperbolic materials, greatly innovates the design principles of traditional imaging lenses.

## 1. Introduction

Einstein's general relativity [1] well predicted gravitational waves [2] and black holes [3] that were recently observed, which has stirred up great interest in exploring and simulating celestial phenomena in diverse physical systems. Transformation optics (TO) [4-6], thanks to its convenience and flexibility on manipulating electromagnetic waves, has been widely used in the analogy of general relativity. TO simulates various models of spacetime by controlling electromagnetic parameters of materials and visualizes the effects by evolution of electromagnetic fields [7]. The effects of some well-known models of spacetime, such as the Schwarzschild black hole [8], the Minkowski spacetime [9], the cosmic string [10], and spatial wormholes [11] have been mimicked by optical materials. Other fantastic topics, such as time travel [12], Milne universe [13], expanding universe [14], were also investigated. Among them, metamaterials play an indispensable role in mimicking the astronomical phenomena, because most of them require complicated anisotropic materials.

The concept of metamaterials is originally derived from negative refractive index that is theoretically studied by Veselago in 1968 [15] and reported by Pendry in 2000 [16]. Among them, negative refraction is thought to be capable of achieving a perfect lens to break the conventional diffraction limit of imaging. Initiated by perfect lens, superlenses [17-20] attracted wide attention owing to their advantages such as light weight, small size, and easy integration, and has been realized in different structures. However, while superlens excels in magnifying evanescent waves and facilitating precise geometric imaging, it is susceptible to the effects of weak material loss [21]. Concurrently, hyperlens [22–25] was proposed as a kind of far-field super-resolution lens because it can convert near-field evanescent waves to far-filed propagating waves and generate a preenlarged image of the input distribution on the output face. Comparing with superlens, hyperlens circumvents the loss effect by having all rays originating from the input plane travel equal path lengths through hyperbolic metamaterials. But it cannot achieve geometrically perfect imaging, which resulting in severe caustics [26]. Hence, designing a geometrically perfect lens without aberration based on hyperbolic system is very important for the development of optical imaging lenses.

Notably, many Lorentzian manifolds [27-28] in Minkowski spacetime are able to provides novel insights and a deeper understanding for key characteristics in hyperbolic systems, which hold great potential for various photoelectric applications, such as imaging. In this paper, we introduce a perfect hyperlens (PHL) inspired by the de Sitter spacetime [29] in cosmology, which can convert evanescent waves to propagating waves and achieve geometrically perfect imaging. Our analysis encompasses the hyperbolic self-focusing and verifies the super-resolution performance of PHL from both geometric optics and wave perspectives. Moreover, we demonstrate our results based on the hyperbolic polaritons by controlling the thickness of α–MoO$_3$ films. We thereby establish a new PHL system through transformation optics, providing innovative guidelines for the future design and optimization of hyperlens.

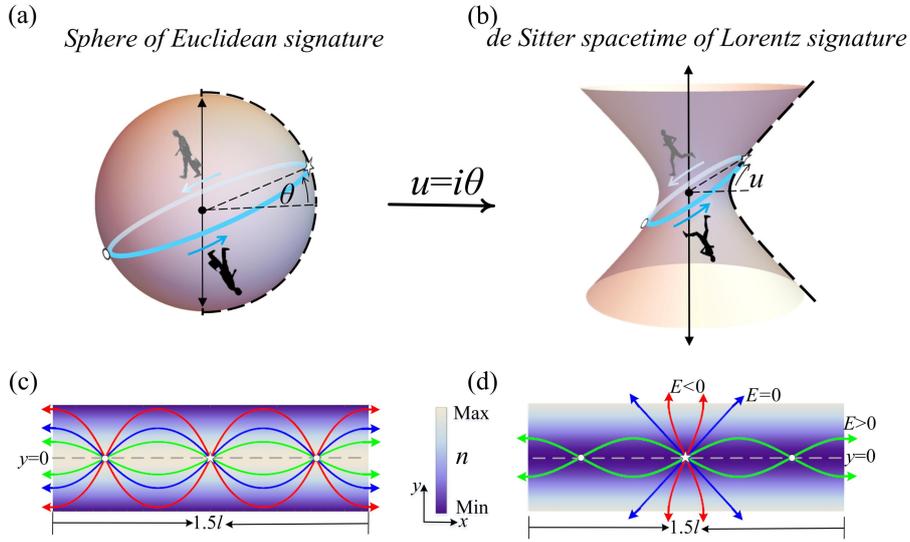

Fig. 1 Schematic of the evolution of the PHL. The "spacetime travel" on the (a) spherical surface of Euclidean signature and (b) the de Sitter spacetime of Lorentz signature. The polar angle $\theta$ varies along the surface of the sphere within a finite range of values [-π/2, +π/2], while $u$ varies along the de Sitter hyperboloid within an infinite range of values [-∞, +∞]. After a Mercator Projection for (a-b), we obtain the (c) ML profile and PHL profile. The stars are the excitation sources ($x_0, y_0$) and the circles are images. In (c-d), the red, blue and green curves represent the rays with incident direction $\left|\frac{dy}{dx}\right|_{initial}$ =0.5/1/2 respectively.

## 2. Results and Discussion

### 2.1 Derivation of PHL profiles

We start from a general static metric with

$$ds^2 = -f(r)dt^2 + \frac{1}{f(r)}dr^2 + r^2 d\Omega^2 , \tag{1}$$

where $t$, $r$ and $\Omega$ represent time, radial coordinate and isolid angle coordinate, respectively. When we set $f(r) = 1 - r^2/a^2$ ($r \leq a$, the horizon), the metric is for de Sitter space. In mathematical physics, de Sitter space is maximally symmetric Lorentzian manifold with constant positive scalar curvature. In addition, it can be defined as a submanifold of generalized Minkowski space of one higher dimension

$$ds^2 = -dx_0^2 + \sum_{i=1}^{n} dx_i^2 . \tag{2}$$

By introducing different slicing method, we can get some special metrics [27]. Here we considering the closed slicing, the metric of the de Sitter space reads

$$ds^2 = -dt^2 + \cosh^2 t \, d\Omega_{n-1}^2. \tag{3}$$

Then we replace $t$ and $\Omega$ with the polar angle $u$ and azimuth angle $\varphi$ respectively, and obtain the new metric represent the de sitter spacetime of Lorentz signature [28]

$$ds^2 = (-du^2 + \cosh^2 u \, d\varphi^2). \tag{4}$$

In Minkowski spacetime, the metric can be sketched as a hyperboloid manifold (see Fig.1 (b)). Interestingly, we can get the spherical metric by set $u=i\theta$ (see Fig. 1(a))

$$ds^2 = (d\theta^2 + \cos^2 \theta \, d\varphi^2). \tag{5}$$

Compared with Eq. (5), Eq. (4) avoids the zeroes of $\cos\theta$ and become asymptotic to de Sitter space in the far past and future. In cosmology, the change from $\theta$ to $u$ describes creation from nothing of a closed universe that then expands exponentially fast. To mimic this case in optics, here we consider the Mercator Projection $z = i\ln(\frac{X+iY}{1-Z})$ [30] for the sphere and de sitter hyperboloid respectively. Then we can get the following metrics in the Cartesian coordinate system

$$ds^2 = n(y)^2(dx^2 + dy^2) = \frac{n_0^2}{\cosh^2(y/a)}(dx^2 + dy^2) \tag{6a}$$

$$ds^2 = n(y')^2(dx^2 - dy'^2) = \frac{n_0^2}{\cos^2(y'/a)}(dx^2 - dy'^2). \tag{6b}$$

Here $ds^2$ also represents the optical path once we treat $n(y)/n(y')$ as refractive index profile, where $a=l/2\pi$ is the lens width across which the on-axis index $n_0$ decrease 1.54 times/increase 1.85 times and $l$ is the path period along the waveguide axis. Equation (6a) is the generalized metric of the 2D Mikaelian lens (ML) profile [31], which has a very interesting property that [32], all rays from a source $(x_0, y_0)$ in the ML will induce images along $y=\pm y_0$ with a period of $a\pi$ (see Fig.1(c)). However, it cannot deal with evanescent waves, resulting in scattered images.

The hyperbolic metric based on Mercator Projection of de sitter hyperboloid Eq. (6b) is the focus of our discussion, which is named as perfect hyperlens (PHL) profile in this paper. Notably, PHL profile can also be obtained by conjugate transformations $y=iy'$ for ML profiles. Crucially, hyperbolic manifold of de Sitter space induces the hyperbolic dispersion and changes the refractive index from a monotone function to a periodic function, but ML profile and PHL profile are still conformally equivalent with perfect geometric imaging and they have the same Gaussian curvature 1 (see Note S2). Furthermore, in addition to Mercator Projection, we also consider the vertical projection for Eq. (4) and obtain other PHL profiles (see Note S4). By various slicing method and mapping methods, we may take more inspiration from cosmological metrics in the future.

*2.2 Self-focusing and imaging analysis of PHL*

We analyze the geometric rays of ML profile and PHL profile with different incident directions $\left|\frac{dy}{dx}\right|_{initial}>1$, $\left|\frac{dy}{dx}\right|_{initial}=1$, $\left|\frac{dy}{dx}\right|_{initial}<1$ (Fig. 1(c-d), Fig. S1). Different from ML, the energy of the PHL system $E$ depends on the incident directions (see Note S1). The rays with $E>0$ can focus perfectly with a period of $a\pi$ while the ray with $E<0$ diverges in the $y$ axis. It's worth noting that, some special rays with $E=0$ will propagate along the line of 45 degrees. If we analyze from the view of cosmology and regard the central lines of PHL profiles $y=jl/2$ (correspond to $n=n_0$) as the center of the universe, the focusing rays and diverging rays can be thought as the motions of the particles moving around the center of the universe under the force of gravity and the particles escaping the center of the universe. Different of monotonic profiles of ML, PHL profiles is periodic with the boundaries at $y=jl/2+l/4$ (correspond to $n=\infty$). Therefore, when we change the incident position to the off-axis, there are exclusive propagation

orbits on each refractive index period (see Fig. S1(b)). As a result, the rays with $E>0$ from the source $(j-1/2)l/2 < y_0 < (j+1/2)l/2$ will induce images along $y=jl-y_0$ with a period of $a\pi$, where $j$ is arbitrary integer.

To consider Eq. (6b) in wave systems, we drop the prime of $y$ for aesthetic reasons and add a $z$-axis as the additional spatial coordinate, and rewrite the line element $dl^2$ and metric $g_{ij}$ as

$$dl^2 = \frac{n_0^2}{\cos^2(y/a)}(dx^2 - dy^2 - dz^2)$$

$$g_{ij} = \begin{bmatrix} \frac{n_0^2}{\cos^2(y/a)} & 0 & 0 \\ 0 & -\frac{n_0^2}{\cos^2(y/a)} & 0 \\ 0 & 0 & -1 \end{bmatrix}. \quad (7)$$

Based on transformation optics, the relative permittivity and relative permeability of the PHL profile can be expressed as [30]

$$\varepsilon^{ij} = \mu^{ij} = -\sqrt{g}\, g^{ij} = \begin{bmatrix} -1 & 0 & 0 \\ 0 & 1 & 0 \\ 0 & 0 & \frac{n_0^2}{\cos^2(y/a)} \end{bmatrix} \quad (8)$$

Since we consider the 2D TE polarization in our study, where the electric field is polarized along $z$-axis ($E_z$), only the parameters $\mu_x$, $\mu_y$ and $\varepsilon_z$ are required. Hence the dominant parameters ($\mu_{xp}$, $\mu_{yp}$, $\varepsilon_{zp}$) can be written as $(-1, 1, n_0^2/\cos(y/a))$ or $(-n_0^2/\cos(y/a), n_0^2/\cos(y/a), 1)$ to remain $n_x^2=\mu_{yp}\varepsilon_{zp}$ and $n_y^2=\mu_{xp}\varepsilon_z$. Then we stimulated the field by the line source along the $z$ direction, which located on the center. Among them, $n_0=1$, $a=1$ and $\lambda=0.9695$ ($k=7$) (arbitary unit). As we know, in the propagation of hyperbolic waves [33], the values of $\mu_x$ and $\mu_y$ control the open angle and direction of hyperbolic waves. When $\mu_x >0$ and $\mu_y <0$, the hyperbolic wave will propagate along the $y$-axis within the open angle $2\arctan(|\mu_y/\mu_x|)$. In our case, the energy will spread out along $y$-axis (see Figs. 2(a-b)). However, when $\mu_x <0$ and $\mu_y >0$, the hyperbolic wave will propagate and image on the $x$-axis periodically (see Figs. 2(c-d)). The wave simulation has high similarity with the ray results. Differently, we can simply choose the appropriate parameters to screen out the imaging waves with $E>0$ because of the high confinement of the hyperbolic waves. Interestingly, the field pattern of PHL profiles is highly similar to the Penrose diagram [34] that used to capture the causal relations between different points in spacetime in compology.

To verify the simulation results theoretically, we solve Maxwell's equations and obtain the electric field solution in Note S3 as

$$E_z(x,y) = \sum_{m=0}^{\infty} c_m \cos^k(y/a) H_2F_1(-m, 2k+m, 1/2+k, (1-\sin(y/a))/2)(\exp(i(k+m)x/a) + \exp(-i(k+m)x/a)). \quad (9)$$

Here coefficient $c_m$ can be expressed as

$$c_m = \int_{-\infty}^{+\infty} \psi_m^*(y) E_z(0,y) dy, \quad (10)$$

where $E_z(0, y)$ is the input electric field at the plane of $x = 0$. For consistency, the input field is set as line current source along $z$ direction here. Notably, the corresponding analytical electric field and intensity are in good agreement with the simulation results (see Figs. 2(e-f)), except for tiny errors caused by computational capacity. Here, our PHL profile shows the excellent focusing effect with ultra-low FWHM about 0.05$\lambda$~0.08$\lambda$ (see Figs. 2(d) and 2(f)). In our opinion, it is because that PHL combines the perfect geometric focusing of the ML with hyperbolic anisotropy that can transfers evanescent wave into propagating wave to support propagating waves with very large wavenumbers [22, 23].

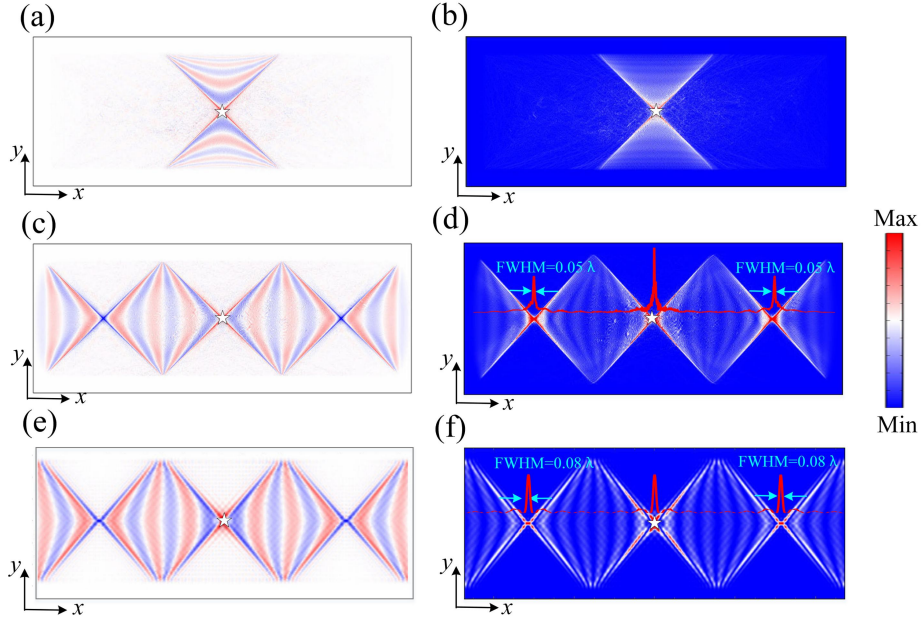

Fig. 2 Light propagation in PHL profile. Numerical simulations of the electric field and the corresponding intensity distributions based on the PHL ($n_0$=1, $a$=1) with (a-b) $\mu_x$ =1, $\mu_y$ =-1 and (c-d) $\mu_x$ =-1, $\mu_y$ =1. Analytical calculations of the electric field and the corresponding intensity distributions based on the PHL with (e-f) $\mu_x$ =-1, $\mu_y$ =1. Here, the stars and red curves in each figure represent the excitation sources at (0,0) and the corresponding FWHM. The wavelength $\lambda$ is set as 0.9695 (arbitary unit).

In order to verify it, we calculate and analyze the corresponding isofrequency contours (IFCs) in Fig. S2. Among them, we discretize the spaces of ML and PHL as infinite interfaces with continuous out-of-plane permittivities $\varepsilon_z$ along $y$ axis. Consider 2D TE modes, the dispersion relation for ML and PHL ($n_0$=$a$=1) can be obtained by

$$\frac{k_x^2}{\mu_y} + \frac{k_y^2}{\mu_x} = \varepsilon_z k_0^2, \qquad (11)$$

where $k_x$, $k_y$ and $k_0$ are the wave vectors in the $x$ and $y$ directions and vacuum wavenumber respectively. Propagating waves refract at the interfaces where the in-plane wave vectors discontinue and such refractions follow the phase-matching conditions ($k_i^\parallel = k_t^\parallel$) [35]. Then we could sketch the incident and refracted wave vectors and determine the propagation path accordingly. In traditional hyperlens [23], all the input distributions with arbitrary wave vectors will be transferred to the output plane through the set of rays parallel to the axis of stratification. Differently, arbitrary input beam with wave vector $k_{in}$=($k_x$, $\pm k_y$) in PHL will be refracted continuously at interfaces with different permittivity $\varepsilon_z$ and finally arrive at the image with $k_{out}$=($k_x$, $\mp k_y$) (red and black isometric arrows in Fig. S2(a)). Comparing isofrequency contours of ML with that of PHL, we can find that PHL can support the propagation and focusing of waves with very large wavenumbers while ML cannot, which will result in significant resolution difference between them. Notably, the beam with higher wave vector can arrive at higher interface with bigger incident angles.

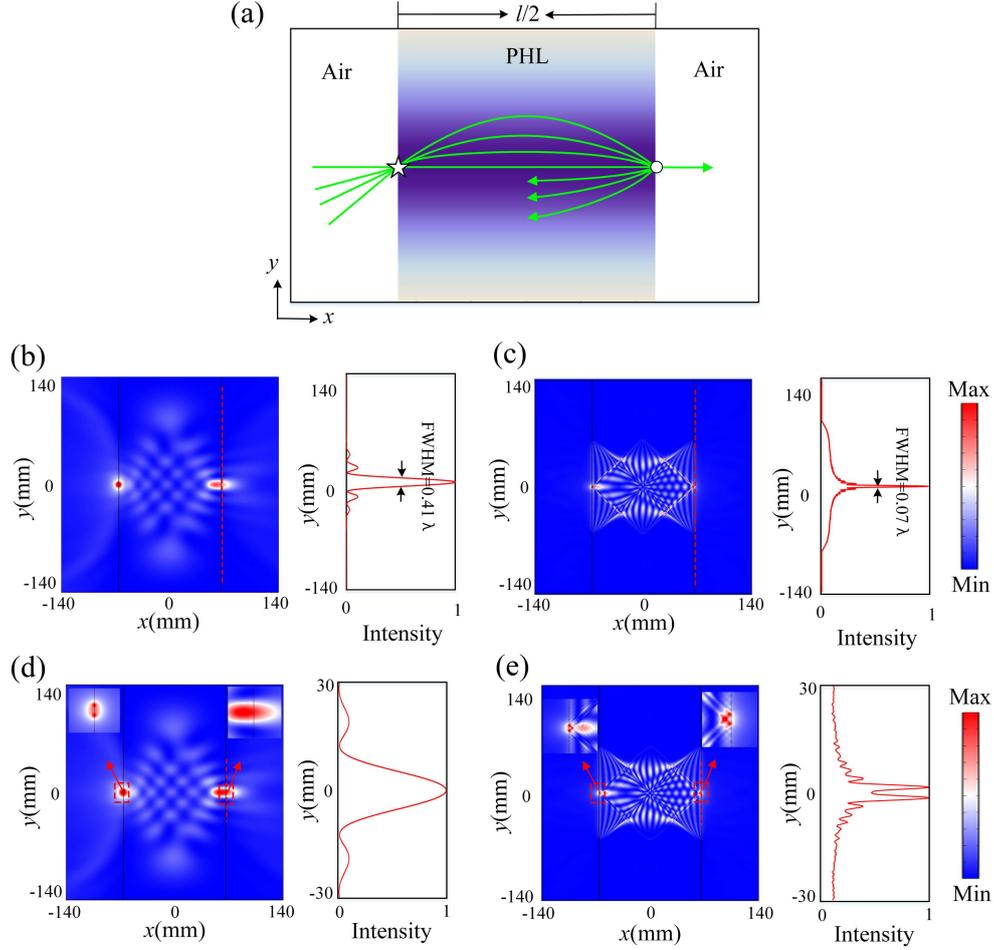

Fig. 3 Comparisons of imaging performance of ML and PHL. (a) The gradient profiles $\sqrt{\varepsilon_z}$ and light ray trajectories of the PHL. Calculated electric field intensity patterns and the corresponding FWHMs in the (b) ML and (c) PHL ($d$=132 mm). The red curves represent the normalized electric field intensity along the $y$ axis direction at the imaging points (red dotted line). The relative FWHMs of the imaging point are marked. Imaging performance of the (d) ML and (e) PHL where two point sources with a spacing of 0.13$\lambda$ are placed at the edge of the lens. The ML fails to resolve the two point sources but PHL can resolve clearly. The simulation frequency is $f$=11GHz.

To further explore the imaging properties, here we consider PHL under the air background (see Fig. 3(a)), whose electromagnetic parameters varies along the $y$ direction as follows:

$$\begin{cases} (\mu_x,\mu_y,\varepsilon_z)=(-1,1,\dfrac{n_0^2}{\cos^2(2\pi y/l)}), & |x|<l/4 \\ (\mu_x,\mu_y,\varepsilon_z)=(1,1,1) & |x|\geq l/4 \end{cases}. \quad (12)$$

It can focus light rays emitted from a point source to another point perfectly, while part of the rays is reflected near the imaging point due to impedance mismatching. In some ways, PHL is similar to the solid immersion lens [36]. Considering flexibility of the sample design in the microwave range, we choose an optimized PHL with $d=l/2=$ 132 mm and $n_0$=1, as an example. In order to reduce the effect of singularity $\varepsilon_z=\infty$ at the boundary $y=\pm l/4$, we added a loss factor of 0.001i to the permeability of PHL. In the simulation, we add the ML for comparison. We put a point source in the air near the lens ($x_0$=-66.05 mm) and numerically calculate the electric field intensity patterns and the corresponding FWHMs with the frequency of $f$=11GHz (see Figures 3(b-c)). The red curves show the electric field intensity and the FWHM of the imaging point is marked on the curve. Comparing with ML, the images in PHL are of higher quality with much smaller spots and lower FWHM. Differently, there are continuous total internal reflection at the boundary of PHL, which causes an energy increasing near the image point. To further confirm the super-resolution imaging of ML and the PHL, a pair of identical point sources with a spacing of 0.13λ are placed on the edge of lens. It is clearly seen that the two identical point sources in PHL are identified (see Fig. 3(e)). In comparison, ML fails to identify two identical point sources, as shown in Fig. 3(d). Therefore, the super-resolution imaging of PHL is demonstrated. More importantly, through coordinate transformation or coordinate substitution, other types of PHL still inherit the characteristics of super-resolution imaging (Fig. S2).

*2.3 Design of the PHL profile based on hyperbolic van der Waals polaritons.*

With these numerical examples demonstrated, we now discuss possible experimental demonstrations of PHL. In order to generate an in-plane hyperbolic response, here we consider two-dimensional hyperbolic van der Waals materials α–MoO$_3$. Phonon polaritons in α-MoO$_3$ [33, 37-38] exhibit in-plane hyperbolic dispersion within most frequencies of the Reststrahlen band (RB) I from 545 to 851 cm$^{-1}$ and Reststrahlen band II from 816 to 972 cm$^{-1}$ (Fig. S3). In addition, highly confined hyperbolic polaritons have the advantages of large sensitivity with the ambient background and the thickness of α–MoO$_3$ flakes. Hence, here we consider the waveguide model composed of three layers: air ($z \geq d$), α–MoO$_3$ slab with finite thickness ($0 \leq z \leq d$), and SiO$_2$ substrate ($z \leq 0$) as shown in Fig. 4(a). Suppose the in-plane PhPs propagate along the [100] direction, i.e., the $x$ axis; then, the electric and magnetic fields can be expressed as $\vec{E}(x,z,t) = \vec{e}E(z)\exp(iqx - i\omega t)$ and $\vec{H}(x,z,t) = \vec{h}H(z)\exp(iqx - i\omega t)$, where the in-plane propagation constant is $q = n_{\text{eff}}k_0$ with $n_{\text{eff}}$ being the effective refractive index and $k_0$ being the vacuum wave vector. Considering the TM modes (with only field components of ($E_y$, $H_x$, and $E_z$)) by solving Maxwell's equations and matching the continuous boundary conditions, we obtain the dispersion relation,

$$k_z d = \arctan(\frac{\alpha_1 \varepsilon_x}{\varepsilon_1 k_z}) + \arctan(\frac{\alpha_3 \varepsilon_x}{\varepsilon_3 k_z}) + m\pi. \qquad (13)$$

where $k_z = \sqrt{k_0^2 \varepsilon_x - \frac{\varepsilon_x q^2}{\varepsilon_z}}$ and $\alpha_{1,3} = \sqrt{q^2 - k_0^2 \varepsilon_{1,3}}$ are $z$ components of photon momenta in α–MoO$_3$, air, and SiO$_2$, respectively. $d$ denotes the thickness of the α–MoO$_3$ slab. $m$=0, 1, 2… are the orders of the TM modes. Taking practical experiments into account, the first order $m$=0 is only considered because the higher modes are usually inhibited by imperfections of the sample edges and the current signal/noise ratio limitation. We use refractive index profile of the PHL [Eq. (6b)] as the approximation of the effective refractive index $n_{\text{eff}}$ of the biaxial slab in the 3D model, which can be obtained by

$$n_{\text{eff}}^2 = \frac{n_0^2}{|\varepsilon_x|\cos^2(y/a)}. \qquad (14)$$

For convenience, we set $a=5/\pi$ ($l/2=5\mu m$). Here, the frequencies 673.5 cm$^{-1}$ with {$\varepsilon_x$(001), $\varepsilon_y$(100), $\varepsilon_z$(010)}={-8.98-0.24i, 8.98-0.06i, 2.87-0.001i} and 935.7 cm$^{-1}$ with {$\varepsilon_x$(100), $\varepsilon_y$(001), $\varepsilon_z$(010)}={-1.36-0.1i, 1.36 -0.02i, 7.52-0.22i} are adopted to excite the hyperbolic response along the $x$ axis dominated by TM polarization and to obtain the nearly 90° opening angle with |Re($\varepsilon_x$)|≈ |Re($\varepsilon_y$)| at the same time. In consideration of the actual thickness of the α–MoO$_3$ flake, we set $n_0/\sqrt{|\varepsilon_x|}$ of the frequencies 673.5 cm$^{-1}$ and 934.7 cm$^{-1}$ as 6.5 and 12 respectively. Substitute $q=n_{\text{eff}}k_0$ into Eq. (13), we

can figure out the relationship between the thickness distribution and the plane coordinates (see Fig.4(b)). In the previous results, we can find that the wave is strictly limited in single period region $-l/4<y<l/4$ because of the singularity at $|y| = l/4$. Hence, we only need to consider and prepare the structure in this region. To simulate the tip-launched polaritons, we introduced a dipole located 200 nm above the top of the α–MoO$_3$ flake and polarized perpendicular to the surface. Consistent with the rays (Fig.1(d)) and two-dimensional wave patterns (Fig. 2(c)) of PHL profiles, the z component of electric fields [Re($E_z$)] on the surface of α–MoO$_3$ films demonstrates the self-focusing responses along the corresponding directions [few scatterings occur in Figs. 4(c-d) due to the limited computational domain and computer capacity, although they are tolerable]. Notably, the focusing period in frequency 935.7 cm$^{-1}$ is less than the estimated value $l/2$=5um while that in frequency 673.5 cm$^{-1}$ is equal to the estimated value. We believe that the error is caused by the high-order effects [39]. Nevertheless, the propagation wavefront and focusing phenomena are still clear. It is worth mentioning that the result can be used to construct hyperbolic multimode waveguide and verify the Talbot Effect in hyperbolic optics [32, 40]. More importantly, the light propagation on the films might be an interesting observation of time travel [12] when regarding y-axis as time axis. In addition, by precisely preparing the films with a gradient thickness profile and the corresponding focusing width, it will be also helpful for infrared super-resolution imaging.

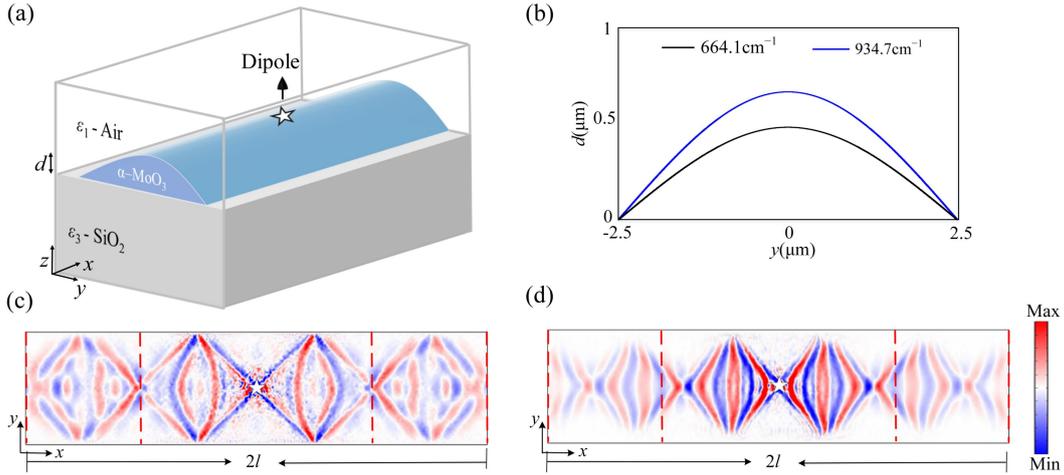

Fig. 4 Design of the PHL profile based on hyperbolic van der Waals polaritons. (a) Schematic of the three-dimensional α–MoO$_3$ waveguide model with gradient thickness. (b) The relation between the thickness of α–MoO$_3$ and the coordinate y. (c) and (d) are polaritonic wave patterns $E_z$ on the top surface of α–MoO$_3$, which are excited by the electric dipoles (stars) of the frequencies 673.5 cm$^{-1}$ and 934.7 cm$^{-1}$, respectively. Among them, the red dotted lines represent the expected images position $x=jl/2$.

## 3. Conclusion

In this paper, we theoretically introduce a perfect hyperlens, which is related to the de Sitter spacetime in cosmology. We validate the excellent super-resolution properties of PHL through theoretical analysis and simulation. Notably, we mainly focus on the simplest case $|\mu_x|=|\mu_y|$ for convenience. In fact, the distinction between the space $|\mu_x|=|\mu_y|$ and space $|\mu_x|\neq|\mu_y|$ lies in the transformation $Y=|\mu_x/\mu_y|y$, rendering both spaces conformal and implying that the ratio of $\mu_x$ and $\mu_y$ only affects the imaging position rather than the performance. Furthermore, it indicates the robustness of PHL to material dispersion and frequencies in structure preparation. In addition to self-focusing application in infrared band, the designed lens is expected to be prepared by means of split-ring resonators, enabling its application in various super-resolution near field imaging systems operating in the microwave band [41-42].

Crucially, our research suggests that some curved spacetime, such as de sitter space, anti-de Sitter space, Schwarzschild black hole could serve as a new basis for the designing of super-resolution lens and give birth to diverse optoelectronics applications except for optical imaging in the future, such as super-scattering and chaos. While our focus lies in optical imaging, the fundamental principle underlying our approach can be extended analogously to diverse wave types, including acoustic [43] and elastic waves [44].

**Funding.** This research was funded by the following: National Key Research and Development Program of China grant 2023YFA1407100 and 2020YFA0710100 (to H.Y.C.); National Natural Science Foundation of China grant 12361161667 (to H.Y.C.); Jiangxi Provincial Natural Science Foundation grant 20224ACB201005 (to H.Y.C.).

**Acknowledgment.** We wish to thank Cheng-Wei Qiu for helpful discussion.

**Competing interests.** The authors declare no competing interests.

**Data availability.** The data that support the findings of this study are all available and archived in the main text or the supplementary materials.

**Supplementary document.** See Supplemental materials for supporting content.

## References

[1] A. Einstein, Approximative integration of the field equations of gravitation. *Sitzungsber. K. Preuss. Akad. Wiss.* **1**, 688(1916).

[2] B. P. Abbott et al., Observation of Gravitational Waves from a Binary Black Hole Merger, *Phys. Rev. Lett.* **116**, 061102 (2016).

[3] K. Akiyama et al. (The event horizon telescope collaboration), First M87 Event Horizon Telescope Results. I. The Shadow of the Supermassive Black Hole. *Astrophys. J. Lett.* **875**, L1 (2019).

[4] J. B. Pendry, D. Schurig, D. R. Smith, Controlling electromagnetic fields. *science* **312**, 1780 (2006).

[5] L. Xu, H. Chen, Conformal transformation optics. *Nat. Photonics* **9**, 15-23 (2015).

[6] L. Xu, H. Chen, Transformation metamaterials. *Adv. Mater.* **33**, 2005489 (2021).

[7] H. Chen, S. Tao, J. Bělín, J. Courtial, R. -X. Miao, Transformation cosmology, *Phys. Rev. A*, **102**, 023528 (2020).

[8] H. Chen, R.-X. Miao, M. Li, Transformation optics that mimics the system outside a Schwarzschild black hole. *Opt. Express* **18**, 15183-15188 (2010).

[9] I. I. Smolyaninov, E. E. Narimanov, Metric signature transitions in optical metamaterials, *Phys. Rev. Lett.* **105**, 067402 (2010).

[10] C. Sheng, H. Liu, H. Chen, S. Zhu, Definite photon deflections of topological defects in metasurfaces and symmetry-breaking phase transitions with material loss. *Nat. Commun.* **9**, 4271 (2018).

[11] A. Greenleaf, Y. Kurylev, M. Lassas, G. Uhlmann, *Phys. Rev. Lett.* **99**, 183901 (2007).

[12] S. R. Boston, Time travel in transformation optics: Metamaterials with closed null geodesics. *Phys. Rev. D* **91**, 124035 (2015).

[13] D. Figueiredo, F. Moraes, S. Fumeron, B. Berche, Cosmology in the laboratory: An analogy between hyperbolic metamaterials and the Milne universe. *Phys. Rev. D* **96**, 105012 (2017).

[14] J. She, S. Tao, T. Liu, H. Chen, Simulation of the expanding universe in hyperbolic metamaterials. *Opt. Express* **31**, 33312-33319 (2023)

[15] V. G. Veselago, The electrodynamics of substances with simultaneously negative values of $\varepsilon$ and $\mu$. *Sov. Phys. -Usp.* **10**, 509(1968).

[16] J. B. Pendry, Negative Refraction Makes a Perfect Lens, *Phys. Rev. Lett*. **85**, 3966-3969 (2000).

[17] N. Fang, H. Lee, C. Sun, X. Zhang, Sub-Diffraction Limited optical imaging with a silver superlens. *Science* **308**, 534 (2005).

[18] T. Taubner, D. Korobkin, Y. Urzhumov, G. Shvets, R. Hillenbrand, Near-Field microscopy through a Sic superlens. *Science* **313**, 1595 (2006).

[19] Y. Xiong, Z. Liu, C. Sun, X. Xiang, Two-dimensional imaging by far-field superlens at visible wavelengths. *Nano Lett.* **7**, 3360–3365 (2007).

[20] X. Zhang, Z. Liu, Superlenses to overcome the diffraction limit. *Nat Mater.* **7**, 435–441(2008).

[21] D. R. Smith, D.Schurig, M. Rosenbluth, S. Schultz, S. A. Ramakrishna, J. B. Pendry,. Limitations on subdiffraction imaging with a negative refractive index slab. *Appl. Phys. Lett.* **82**, 1506-1508 (2003).

[22] Z. Jacob, L. V. Alekseyev, E. Narimanov, Optical Hyperlens: Far-field imaging beyond the diffraction limit. *Opt. Express* **14**, 8247-8256 (2006).

[23] A. Salandrino, N. Engheta, Far-field subdiffraction optical microscopy using metamaterial crystals: Theory and simulations. *Phys. Rev. B* **74**, 075103 (2006).

[24] J. Li, L. Fok, X. Yin, G. Bartal, X. Zhang, Experimental demonstration of an acoustic magnifying hyperlens. *Nature Mater.* **8**, 931–934 (2009).

[25] J. Rho, Z. L. Ye, Y. Xiong, X. B. Yin, Z. W. Liu, H. Choi, G. Bartal and X. Zhang, Spherical hyperlens for two-dimensional sub-diffractional imaging at visible frequencies. *Nat. Commun.* **1**, 143 (2010).

[26] D. R. Smith, D.Schurig, M. Rosenbluth, S. Schultz, S. A. Ramakrishna, J. B. Pendry,. Limitations on subdiffraction imaging with a negative refractive index slab. *Appl. Phys. Lett.* **82**, 1506-1508 (2003).

[27] W. de Sitter, On the curvature of space. *Proc. Kon. Ned. Acad. Wet.*, **20**, 229–243(1917).

[28] E. Witten, A note on complex spacetime metrics. *Frank Wilczek: 50 years of theoretical physics*. 245-280 (2022).

[29] W. de Sitter, On the relativity of inertia: Remarks concerning Einstein's latest hypothesis. *Proc. Kon. Ned. Acad. Wet.* **19**, 1217–1225,

[30] U. Leonhardt, T. Philbin, *Geometry and light: the science of invisibility*. (Courier Corporation, 2012).

[31] A. Mikaelian, A. Prokhorov, V self-focusing media with variable index of refraction. *Prog. Optics* **17**, 279-345 (1980).

[32] X. Wang, H. Y. Chen, H. Liu, L. Xu, C. Sheng, S. Zhu, Self-focusing and the Talbot effect in conformal transformation optics. *Phys. Rev. Lett.* **119**, 033902 (2017).

[33] G. Hu, Q. Ou, G. Si, Y. Wu, J. Wu, Z. Dai, A. Krasnok, Y. Mazor, Q. Zhang, Q. Bao, C. W. Qiu, A. Alù, Topological polaritons and photonic magic angles in twisted α-MoO$_3$ bilayers. *Nature* **582**, 209-213 (2020).

[34] G. F. R. Ellis, S. W. Hawking, The large scale structure of space-time. Cambridge university press, 1973.

[35] J. A. Kong, Electromagnetic Wave Theory. John Wiley & Sons: New York, 1990.

[36] Y. Zhou, Z. Hao, P. Zhao, H. Chen, Solid immersion maxwell's fish-eye lens Without drain. *Phys. Rev. Appl.* **17**, 034039 (2022).

[37] W. Ma, P. Alonso-González, S. Li, A. Y. Nikitin, J. Yuan, J. Martín-Sánchez, J. Taboada Gutiérrez, I. Amenabar, P. Li, S. Vélez, C. Tollan, Z. Dai, Y. Zhang, S. Sriram, K. Kalantar-Zadeh, S.-T. Lee, R. Hillenbrand, Q. Bao, In-plane anisotropic and ultra-low loss polaritons in a natural van der Waals crystal. *Nature* **562**, 557–562 (2018).

[38] Z. Zheng, N. Xu, S. L. Oscurato, M. Tamagnone, F. Sun, Y. Jiang, Y. Ke, J. Chen, W. Huang, W. L. Wilson, A. Ambrosio, S. Deng, H. Chen, A mid-infrared biaxial hyperbolic van der Waals crystal. *Sci. Adv.* **5**, eaav8690 (2019).

[39] G. Álvarez-Pérez, K. V. Voronin, V. S. Volkov, P. Alonso-González, A. Y. Nikitin, Analytical approximations for the dispersion of electromagnetic modes in slabs of biaxial crystals. *Phys. Rev. B* **100**, 235408 (2019)

[40] S. Li, Y. Zhou, J. Dong, et al, Universal multimode waveguide crossing based on transformation optics. *Optica* **5**, 1549-1556 (2018).

[41] D. Schurig, J. J. Mock, B. J. Justice, S. A. Cummer, J. B. Pendry, A. F. Starr, D. R. Smith. Metamaterial Electromagnetic Cloak at Microwave Frequencies. *Science* **314**, 977-980 (2006).

[42] Y. G. Ma, C. K. Ong, T. Tyc, U. Leonhardt. An omnidirectional retroreflector based on the transmutation of dielectric singularities. *Nat. Mater.* **8**, 639-642 (2009).

[43] L. Quan, A. Alù. Hyperbolic sound propagation over nonlocal acoustic metasurfaces. *Phys. Rev. Lett.* **123**, 244303 (2019).

[44] D. Lee, Y. Hao, J. Park, I. S. Kang, S.-H. Kim, J. Li, J. Rho, Singular lenses for flexural waves on elastic thin curved plates, *Phys. Rev. Appl.* **15**, 034039 (2021).

# Supplementary Materials for

# "Cosmology analogy for perfect hyperlens"

Tao Hou[1], Wen Xiao[1], Huanyang Chen[1,2*]

[1]Department of Physics, Xiamen University; Xiamen, 361005, China
[2]Jiujiang Research Institute of Xiamen University; Jiujiang, 332000, China
*kenyon@xmu.edu.cn

**Note S1:** The calculation and analysis of the rays.

**Note S2:** Curvature analysis.

**Note S3:** Wave analysis.

**Note S4:** Other PHLs and their transformation connection.

**Note S5:** Optical parameters of natural van der Waals crystal α-$MoO_3$.

**Figs. S1 - S4**

**Supplementary References**

**Note S1: The calculation and analysis of the rays.**

In order to analyze the rays on the plane with the non-Euclidean metric, we consider the two-dimensional geodesic equation:

$$\frac{d^2 x^\lambda}{d\xi^2} + \Gamma^\lambda_{uv} \frac{dx^u}{d\xi}\frac{dx^v}{d\xi} = 0 \tag{S1}$$

where $\{\lambda, u, v\}$ are dummy indexes (only with values 1 or 2), $\Gamma^\lambda_{uv} = g^{\lambda\omega}(\partial_v g_{u\omega} + \partial_u g_{v\omega} - \partial_\omega g_{uv})/2$ are Christoffel symbols of the second kind in Riemannian physical space, and $\xi$ is the length parameter of the geodesics. Consider the line element of PHL profile $ds^2 = n(y)^2(dx^2 - dy^2) = \frac{n_0^2}{\cos^2(y/a)}(dx^2 - dy^2)$, we get

$$\frac{d^2 x}{d\xi^2} + \frac{2}{a}\tan(y/a)\frac{dx}{d\xi}\frac{dy}{d\xi} = 0 \tag{S2a}$$

$$\frac{d^2 y}{d\xi^2} + \frac{1}{a}\tan(y/a)\left(\left(\frac{dx}{d\xi}\right)^2 + \left(\frac{dy}{d\xi}\right)^2\right) = 0 \tag{S2b}$$

By solving the Eq. (S2a-S2b), we can obtain the geodesics of PHL profile numerically. Further, we carry out the analytical solutions. It is not difficult to find that Eq. (S2a) can be solved as

$$n(y)^2 \frac{dx}{d\xi} = L. \tag{S3}$$

Here, the arbitrary integral constants $L$ is the momentum of the system. It is worth mentioning that for any free trajectory (i.e., without external force, collision, etc), the quantity $n(y)^2 \, dx/d\xi$ remains invariant once the initial position and direction, are given. With Eq. (S3), Eq. (S2b) can be reduced to

$$-\frac{n(y)^2}{2}\left(\frac{dy}{d\xi}\right)^2 + \frac{L^2}{2n(y)^2} = E \tag{S4}$$

where $E$ is the total energy of the system. Combining Eq. (S3) and Eq. (S4), we can obtain that

$$\frac{dy}{dx} = \pm\frac{\sqrt{L^2 - 2En(y)^2}}{L} = \pm\frac{\sqrt{L^2 - 2En_0^2/\cos^2(y/a)}}{L}, \tag{S5}$$

where $\pm$ depends on the sign of $\left[\frac{dy}{dx}\right]_{initial}$. It is not difficult to find that $\left|\frac{dy}{dx}\right|_{initial} < 1$ is corresponding to the case of $E>0$ while $\left|\frac{dy}{dx}\right|_{initial} > 1$ is corresponding to the case of $E<0$. In the main text, we show the rays on the axis $y=0$ of PHL. Influenced by the incident direction, the rays will diverge or focus on the axis. Here, we change the incident position to the off-axis, the rays of ML profile and PHL profile are shown in the Fig. S1.

Different from ML, the refractive index profiles of PHL is periodic. Therefore, there are exclusive propagation orbits on each refractive index period (see Fig. S1(b)). In addition, we can find the green/blue/red rays keep the same characteristic as Fig. 1(d), which verify the relationship between system energy $E$ and $\left|\frac{dy}{dx}\right|_{initial}$.

**Note S2: Curvature analysis.**

For arbitrary surface in $\mathbf{R}^3$, Gaussian curvature [S1] can be expressed as the ratio of the determinants of the second fundamental forms II and the first fundamental forms I:

$$K = \frac{\det(\text{II})}{\det(\text{I})} = \frac{h_{11}h_{22} - h_{12}^2}{g_{11}g_{22} - g_{12}^2}, \tag{S6}$$

where $\begin{bmatrix} g_{11} & g_{12} \\ g_{12} & g_{22} \end{bmatrix}$ and $\begin{bmatrix} h_{11} & h_{12} \\ h_{12} & h_{22} \end{bmatrix}$ are coefficients of the second and the first fundamental forms. For an orthogonal parametrization ($g_{12} = 0$), the Gaussian curvature in the Cartesian coordinates can be written in terms of the first fundamental form as

$$K = -\frac{1}{2\sqrt{g_{11}g_{22}}} \left( \frac{\partial}{\partial x} \frac{\frac{\partial g_{22}}{\partial x}}{\sqrt{g_{11}g_{22}}} + \frac{\partial}{\partial y} \frac{\frac{\partial g_{11}}{\partial y}}{\sqrt{g_{11}g_{22}}} \right). \tag{S7}$$

In 2D, the Ricci scalar curvature $R_{\text{ric}}$ relates to the Gaussian curvature $K$ through $R_{\text{ric}} = 2K$. It is worth mentioning that some traditional optical lenses yield constant Gaussian curvatures. For example, $K(\text{Maxwell Fish-eye lens}) = K(\text{ML}) = 1$ and $K(\text{Poincaré disk}) = -1$. Here, we set

$$dl^2 = \frac{n_0^2}{\cosh^2(\alpha y / a)}(dx^2 + \alpha^2 dy^2) \begin{cases} \alpha = 1 & e \\ \alpha = i \text{ Perfect Hyperlens} \end{cases}. \tag{S8}$$

According to Eqs. (S7) and (S8), the Gaussian curvature $K$ remains 1 regardless of the value of $\alpha$. From another point of view, they are conformal equivalent with perfect geometric imaging.

**Note S3: Wave analysis.**

Since we consider the 2D TE polarization in our study, where the electric field is polarized along $z$-axis ($E_z$), only the parameters $\mu_x$, $\mu_y$ and $\varepsilon_z$ are required. Now the electric field equation can be reduced and written as

$$\mu_y \frac{\partial^2 E_z}{\partial x^2} + \mu_x \frac{\partial^2 E_z}{\partial y^2} + \varepsilon_z(y) k_0^2 E_z = 0, \tag{S9}$$

where $k_0$ is the wave vector. By separating the variables, we assume the solution as the form $E_z(x,y) = \sum_{m=0}^{\infty} c_m \psi_m(y) \exp(i k_x x/a)$, where $k_x$ is the constant of propagation along the $x$-axis and $c_m$ are coefficients. With Eqs. (8) and (S9), we obtain the following equation for $\psi_m(y)$

$$-\frac{\partial^2 \psi_m}{\partial y^2} + \left(\frac{n_0^2 k_0^2}{\cos^2(y/a)} - k_x^2\right)\psi_m = 0. \tag{S10}$$

Unlike the wave solution of ML, the wave solution of THL cannot be expressed in terms of associated Legendre polynomial because it doesn't support complex arguments. Therefore, here we set the solutions of Eq. (S10) as

$$\psi_m(y) = \mathrm{Cos}^k(y/a) H_2F_1(k - k_x, k + k_x, 1/2 + k, (1 - \mathrm{Sin}(y/a))/2) \tag{S11}$$

where $k = \frac{1 + \sqrt{4a^2 n_0^2 k_0^2 + 1}}{2}$. To fit the hyperbolic dispersion $|k_x| > k_0$, we set $k_x = \pm(k+m)$, $m = 0, 1, 2, \ldots$
In this case, the electric field solution will be

$$E_z(x,y) = \sum_{m=0}^{\infty} c_m \mathrm{Cos}^k(y/a) H_2F_1(-m, 2k+m, 1/2+k, (1-\mathrm{Sin}(y/a))/2)(\exp(i(k+m)x/a) + \exp(-i(k+m)x/a)) \tag{S12}$$

From (S12) it follows that at $x = 2\pi a p$, $p = 1, 2, 3\ldots$, the module of the complex amplitude will be periodically repeated. It is worth noting that, the value of function $\psi_m(0)$ remains 0 when $m$ is odd. In other words, the wave propagating on the $x$ axis has a periodicity of $a\pi$. Indeed, the intensity of the electric field Eq. (S12) is

$$I(x, y=0) = |E_z(x, y=0)|^2 = \sum_{m=0}^{\infty} c_m^2 \psi_m^2(0) \pm 2 \sum_{m_1 < m_2} c_{m_1} c_{m_2} \psi_{m_1}(0)\psi_{m_2}(0)(\cos((m_1 - m_2)x/a) + \cos((2k + m_1 + m_2)x/a)). \tag{S13}$$

The derivative of the intensity is

$$\frac{dI(x, y=0)}{dx} = \pm 2 \sum_{m_1 < m_2} c_{m_1} c_{m_2} \psi_{m_1}(0)\psi_{m_2}(0)\left(\frac{(m_1 - m_2)}{a}\sin((m_1 - m_2)x/a) + \frac{(2k + m_1 + m_2)}{a}\sin((2k + m_1 + m_2)x/a)\right). \tag{S14}$$

From (S14) it follows that this derivative is zero at $x = \pi a p/2$, i.e. there will be intensity maxima and minima at these points of the optical axis $y=0$. Because at $x = \pi a p$ the intensity is reproduced, at the intermediate points $x = \pi a/2 + \pi a p$ the light field will be either collimated (intensity minimum) or focused (intensity maximum). We can find that the propagation law is the same as that of the ML [S2]. Differently, the evanescent part ($k_y \gg k_0$) will decay in ML but propagate in the PHL.

**Note S4: Other PHLs and their transformation connection.**

In addition to Mercator Projection, we also consider the vertical projection for Eq. (4), i.e. $u \rightarrow y$ and $\varphi \rightarrow x$. Then we obtain a new profile and consider its general form (call it 2-PHL here)

$$ds^2 = n_0^2(-dy^2 + \cosh^2(y/a)dx^2) \ . \tag{S15}$$

Transformation optics can relate the different refractive-index profiles of optical lenses by coordinate transformations. For example, Maxwell fish-eye lens can be transformed into ML and Eaton lens by conformal mapping $w=e^z$ [30] and inverse function transformation [S3] respectively. In the transformation optics theory, after a two-dimensional coordinate transformation ($z_p = z_v$), $x_p = x_p(x_v, y_v)$ and $y_p = y_p(x_v, y_v)$, Maxwell's equations keep their forms for coordinates ($x_p, y_p, z_p$). The connection of electromagnetic parameters between physical space ($x_p, y_p, z_p$) and virtual space ($x_v, y_v, z_v$) are written as

$$\hat{\varepsilon}_p = \frac{\hat{J}\hat{\varepsilon}_v \hat{J}^T}{|\det(\hat{J})|} \ , \ \hat{\mu}_p = \frac{\hat{J}\hat{\mu}_v \hat{J}^T}{|\det(\hat{J})|}, \tag{S16}$$

where $\hat{J}$ is the Jacobian transformation matrix:

$$\hat{J} = \begin{bmatrix} \dfrac{\partial x_p}{\partial x_v} & \dfrac{\partial x_p}{\partial y_v} & 0 \\ \dfrac{\partial y_p}{\partial x_v} & \dfrac{\partial y_p}{\partial y_v} & 0 \\ 0 & 0 & 1 \end{bmatrix} \tag{S17}$$

In order to demonstrate the transformation relations between PHL profiles and 2-PHL profiles, here we consider a transformation $x_p = x_v$, $\tanh(y_p/2a) = \tan(y_v/2a)$, i.e. $y_p = 2a\mathrm{atanh}(\tan(y_v/2))$, and set electromagnetic parameters of virtual space as PHL profile. Based on (S16) and (S17). We get the electromagnetic parameters

$$\hat{\varepsilon}_p = \hat{\mu}_p = \begin{bmatrix} -\mathrm{sech}(y_p/a) & 0 & 0 \\ 0 & \cosh(y_p/a) & 0 \\ 0 & 0 & n_0^2 \cosh(y_p/a) \end{bmatrix}. \tag{S18}$$

According to Eqs. (8) and (S18), we can obtain the same metric as 2-PHL. At this point, we have proved that 2-PHL profiles is conformally equivalent to PHL profiles.

Since we consider the 2D TE polarization in our study, the dominant parameters can be written as $(\mu_{xp}, \mu_{yp}, \varepsilon_{zp})=(-1, \cosh^2(y_p/a), n_0^2)$ to keep $n_{xp}^2=\mu_{yp}\varepsilon_{zp}$ and $n_{yp}^2=\mu_{xp}\varepsilon_{zp}$. Using the same settings as PHL, we calculate the rays and the imaging performance of the 2-PHL ($a=n_0=1$), as shown in the Fig. S3. We can find that the rays with $\left|\dfrac{dy}{dx}\right|_{initial}=0.5/2$ still focus/diverge but $\left|\dfrac{dy}{dx}\right|_{initial}=1$ is not critical position because of the influence of the coordinate transformation. Comparing with the PHL, the wavefront of 2-PHL is wider than that of PHL because the transformation $y_p = 2a\mathrm{atanh}(\tan(y_v/2a)) > y_v$, which results in weaker interfacial reflection. Of course, the 2-PHL also has a good focusing and super-resolution effect.

In addition to 2-PHL, through different mapping methods, we can also design more types of PHL, which will provide important references for the design of future super-resolution imaging devices. At this point, we have established a new PHL system through transformation optics.

**Note S5: Optical parameters of natural van der Waals crystal α-MoO₃.**

α-MoO₃ is a natural anisotropic van der Waals crystal and its permittivity can be modeled by the Lorentz model [33]

$$\varepsilon_j = \varepsilon_\infty^j \left(1 + \frac{\omega_{LO}^{j\,2} - \omega_{TO}^{j\,2}}{\omega_{TO}^{j\,2} - \omega^2 - i\omega\gamma^j}\right), \quad j = x, y, z \tag{S19}$$

The parameters $\varepsilon_\infty^j$ are the high-frequency dielectric constant along the direction $j$. Parameters $\omega_{LO}^{j\,2}$ and $\omega_{TO}^j$ are the longitude and transverse optical phonon resonance frequencies, respectively. $\gamma^j$ is inelastic loss rates of the material. The coordinates $x$, $y$, and $z$ are oriented along the three principal axes of the crystal, which correspond to the crystallographic directions [100], [001], and [010] of α-MoO₃, respectively. The relation between real parts of the permittivity of α–MoO₃ and frequency is illustrated in Fig. S4, and the corresponding parameters are listed in Table S1.

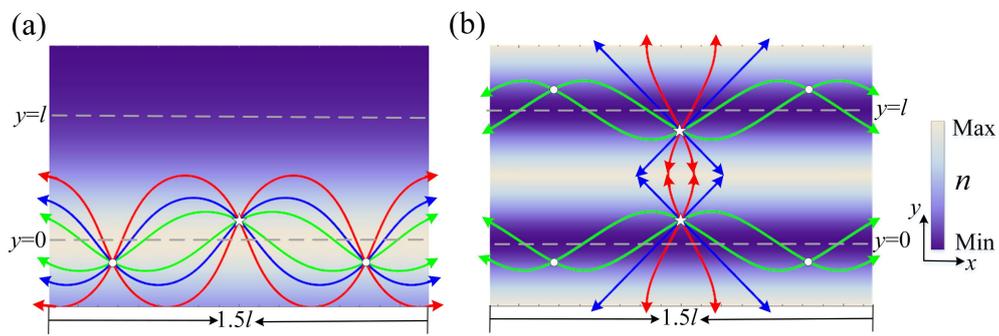

Fig. S1 The rays in ML and PHL from the off-axis sources. (a)The rays from the sources (0,0.5) in the ML. The rays from the sources (b) (0,0.5) and (0, π-0.5) in the PHL. The red, blue and green curves represent the rays with incident direction $\left|\frac{dy}{dx}\right|_{initial}$ =0.5/1/2 respectively. The stars are the excitation sources and the circles are images.

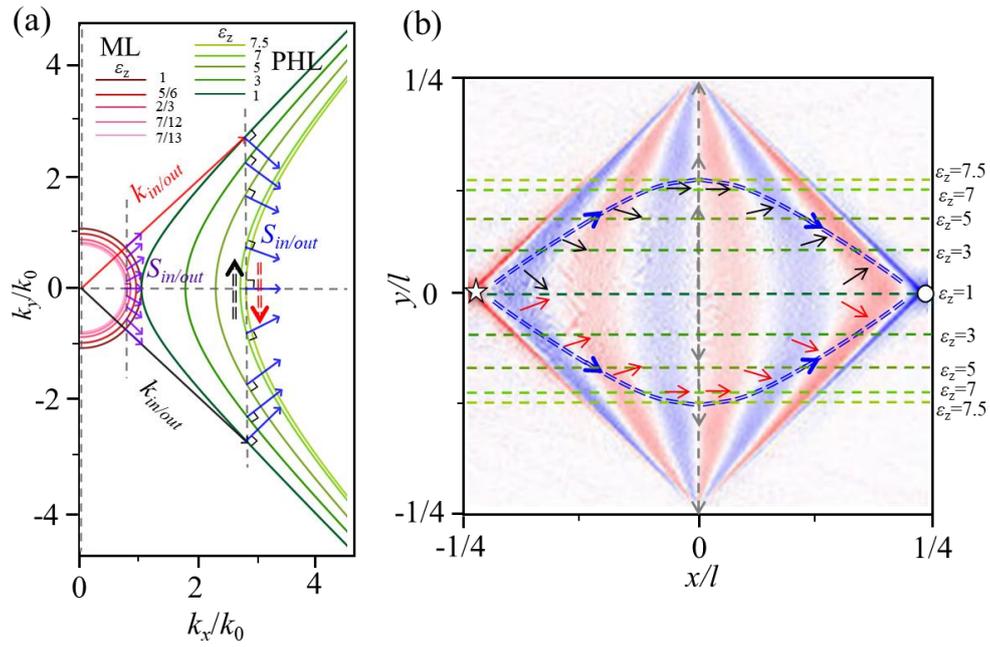

Fig. S2 IFC analysis of ML and PHL profiles. (a) Momentum conservation on IFCs of ML (cicle) and PHL (hyperbola). Here we only show part of IFCs with different out-plane permittivity $\varepsilon_z$. According to phase-matching conditions, we sketch the Poynting vectors (blue and purple arrows) on the IFCs. (b) Field distribution of PHL during single focusing period. In vector space (a), the red isometric arrows and black isometric arrows represent the evolution of the wave vector $k$ from the top to the bottom and from the bottom to the top respectively. In real space (b), they represent the lower light path and the upper light path.

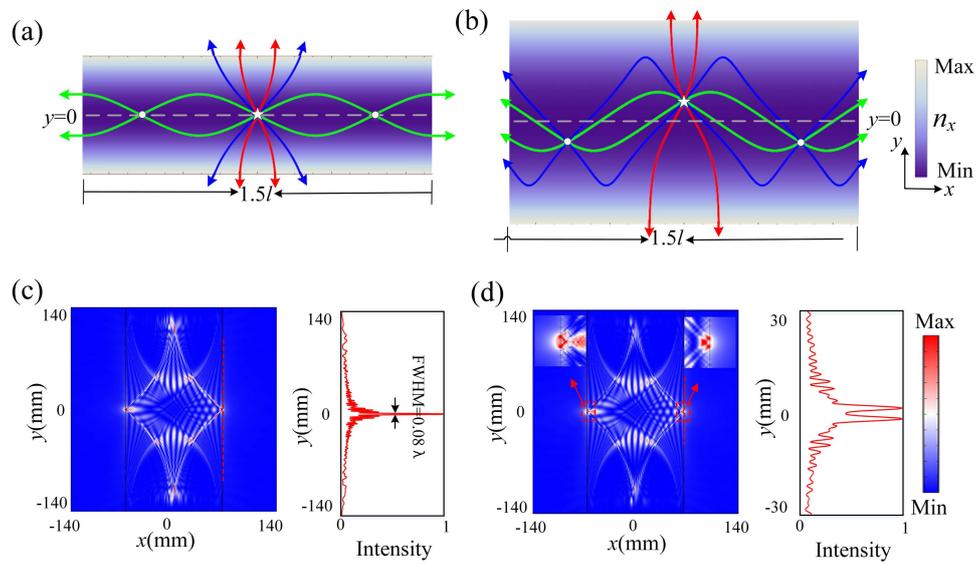

Fig. S3 Propagation and imaging performance of 2-PHL. The rays from the sources (a) (0,0) and (b) (0,0.5) in the 2-PHL. The red, blue and green lines represent the rays with incident direction $\left|\frac{dy}{dx}\right|_{initial}$ =0.5/1/2 respectively. The stars are the excitation sources and the circles are images. (c) Calculated electric field intensity patterns and the corresponding FWHMs in the 2-PHL. The red curves represent the normalized electric field intensity along the $y$ axis direction at the imaging point. The relative FWHMs of the imaging point are marked. (d) Imaging performance of the 2-PHL in which two point sources with a spacing of 0.15λ are placed at the edge of the lens. 2-PHL can resolve clearly. The simulation frequency in the figures is $f$=11GHz.

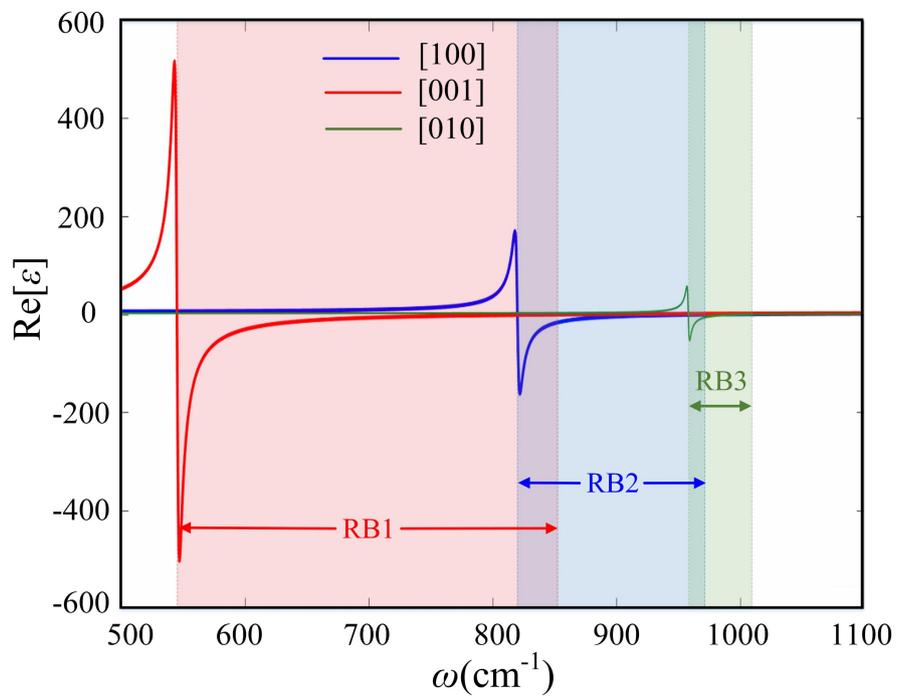

Fig. S4 Real-part permittivities of α-MoO$_3$ along different principal directions. The permittivities are obtained by fitting the obtained data with a Lorentzian model. Three different Reststrahlen bands of α–MoO$_3$ are shaded in different colors.

Table S1 Parameters used for modeling the permittivity of α-MoO$_3$.

| Direction | $\varepsilon_\infty$ | $\omega_{LO}(cm^{-1})$ | $\omega_{TO}(cm^{-1})$ | $\gamma(cm^{-1})$ |
|---|---|---|---|---|
| x[100] | 4 | 972 | 820 | 4 |
| y[100] | 5.2 | 851 | 545 | 4 |
| z[100] | 2.4 | 1004 | 958 | 2 |

**Supplementary References**

[S1]. S. M. Carroll, Spacetime and Geometry (Cambridge University Press, 2019).

[S2]. V. V. Kotlyar, A. A. Kovalev, V. A. Soifer, Subwavelength focusing with a Mikaelian planar lens. Opt. Memory Neural 19, 273-278 (2010).

[S3]. J. C. Miñano, P. Benítez, A. Santamaría. Hamilton-Jacobi equation in momentum space. Opt. Express 14, 9083-9092 (2006).